\documentclass[11pt]{article}
\usepackage{moriond,epsfig}

\usepackage{amsmath}

\usepackage{graphicx}

\usepackage{paralist}

\def\Journal#1#2#3#4{{#1} {\bf #2}, #3 (#4)}

\def\slantfrac#1#2{\kern.1em^{#1}\kern-.3em/\kern-.1em_{#2}}

\begin{document}

\vspace*{4cm}
\title{I.C.E.: AN ULTRA-COLD ATOM 
SOURCE FOR LONG-BASELINE INTERFEROMETRIC INERTIAL SENSORS IN REDUCED GRAVITY}

\def\LCFIO{\unskip$^{\dagger}$}
\def\ONERA{\unskip$^{\ddagger}$}
\def\SYRTE{\unskip$^{\star}$}
\def\CNES{\unskip$^{\mathchar "278}$}

\author{%
    \mbox{G. VAROQUAUX\LCFIO,}
    \mbox{N. ZAHZAM\ONERA,}
    \mbox{W. CHAIBI\SYRTE,}
    \mbox{J-F. CL\'EMENT\LCFIO,}
    \mbox{O. CARRAZ\ONERA,}
    \mbox{J-P. BRANTUT\LCFIO,}
    \mbox{R. A. NYMAN\LCFIO,}
    \mbox{F. PEREIRA DOS SANTOS\SYRTE,}
    \mbox{L. MONDIN\CNES},
    \mbox{M. ROUZE\CNES}, 
    \mbox{Y. BIDEL\ONERA,}
    \mbox{A. BRESSON\ONERA,}
    \mbox{A. LANDRAGIN\SYRTE,} and
    \mbox{P. BOUYER\LCFIO\footnotemark}
}

\footnotetext{philippe.bouyer@institutopique.fr \quad --- \quad
 \texttt http://www.ice-space.fr}

\address{$\dagger$ Laboratoire Charles Fabry de l'Institut d'Optique,
Campus Polytechnique, RD 128, 91127 Palaiseau, France}

\address{$\ddagger$ Office National d'\'Etude et de Recherches
A\'erospatiales, Chemin de la Huni\`ere, 91761 Palaiseau, France}

\address{$\star$ LNE-SYRTE, UMR8630, Observatoire de Paris,
61 avenue de l'Observatoire, 75014 Paris, France}

\address{$\mathchar "278$ CNES DCT/SI/OP, 18, Avenue Edouard Belin
31401 Toulouse CEDEX 9, France}

\maketitle\abstracts{The accuracy and precision of current
atom-interferometric inertial sensors rival state-of-the-art conventional
devices using artifact-based test masses\cite{Peters:2001}. Atomic
sensors are well suited for fundamental measurements of gravito-inertial
fields. The sensitivity required to test gravitational theories can be
achieved by extending the baseline of the
interferometer\cite{Dimopoulos:2007}. The I.C.E. ({\sl
Interf\'erom\'etrie Coh\'erente pour l'Espace}) interferometer aims to
achieve long interrogation times in compact apparatus via reduced
gravity. We have tested a cold-atom source during airplane parabolic
flights. We show that this environment is compatible with free-fall
interferometric measurements using up to 4 second interrogation time.
We present the next-generation apparatus using degenerate gases for low
release-velocity atomic sources in space-borne experiments. }


Inertial sensors are useful devices in both science and industry. Higher
precision sensors could find scientific applications in the areas of
general relativity\cite{chow}, navigation, surveying and analysis of
Earth structures. Matter-wave interferometry was first envisaged to probe
inertial forces\cite{clauser}. Neutron interferometers were used to
measure the acceleration due to gravity\cite{werner} and the rotation of
the Earth\cite{colella} at the end of the 1970s. In 1991, atom
interference techniques were used in proof-of-principle work to measure
rotations\cite{Borde91} and accelerations\cite{chu}. Many theoretical and
experimental works have been performed to investigate this new kind of
inertial sensors\cite{Bermann:1997}. Some of the recent works have since
shown very promising results leading to a sensitivity comparable to other
kinds of sensors, for both rotation\cite{Gustavson97,todd,Canuel} as for
acceleration\cite{Peters:2001,Peters97}.

\section{Atoms in microgravity as probes of the gravito-inertial field}

\subsection{Using atoms as test masses.}

Following pioneering work on atomic clocks \cite{Laurent:2006},
ultra-high precision inertial sensors are expected to be used in
micro-gravity for tests of gravitation theories or to provide accurate
and exact drag-free motion that is required for deep-space mapping of
gravity\cite{PioneerAnomaly}. Closer to Earth, they can lead to possible
experiments that could test the Einstein equivalence
principle\cite{Lammerzahl:2006}: Lorentz invariances, the universalities
of a free fall and gravitational redshift, as well as the constancy of
gravitational and fine-structure constants or higher order gravitational
effects such as the Lense-Thirring effect\cite{GPB,HYPER}.

Conventional gravity and acceleration probes\cite{GPB} rely on
artifact-based macroscopic test masses to probe the gravito-inertial
field. Using atoms as proof masses directly relates measurements to
fundamental quantities, without the need for geometrical factors. The test
masses are not subject to manufacturing errors and their displacement is
referenced to the well-controlled wavelength of a pair of laser beams,
providing long term accuracy and precision.

\subsection{Atom interferometry and precision gravimetry}

An atom interferometer measures the phase-shift acquired by atoms through
different coherence-preserving paths. Since the phase acquired by an atom during its free propagation is strongly
dependent on the gravito-inertial field it experiences\cite{Borde:2001}, the
interferometric read-out of this phase-shift can give access to direct
measurements of the metric tensor\cite{Borde:2000}.

In most atom-interferometry
experiments, an ensemble of particles is split into two different paths
by a coherent beam-splitting process\cite{Bermann:1997}. After a phase
accumulation time $T$, the two paths are recombined by a second
beam-splitting process. The probabilities of detecting particles in the
two output channels of the beam-splitter are given by
quantum interference of the two coherent propagation paths and are
sinusoidal functions of the accumulated phase difference. The ultimate
precision in the read-out of the phase is limited by the number of
detected particles and scales as $\Delta\phi_\text{min} = 2\pi/\sqrt{N}$
(quantum projection noise limit\cite{Wineland:1992}). Typically
$\Delta\phi_\text{min} \sim 2\,\text{mrad}$ for $10^{7}$ particles.

\begin{figure}[b]
\begin{center}
\vspace*{-1ex}
\includegraphics{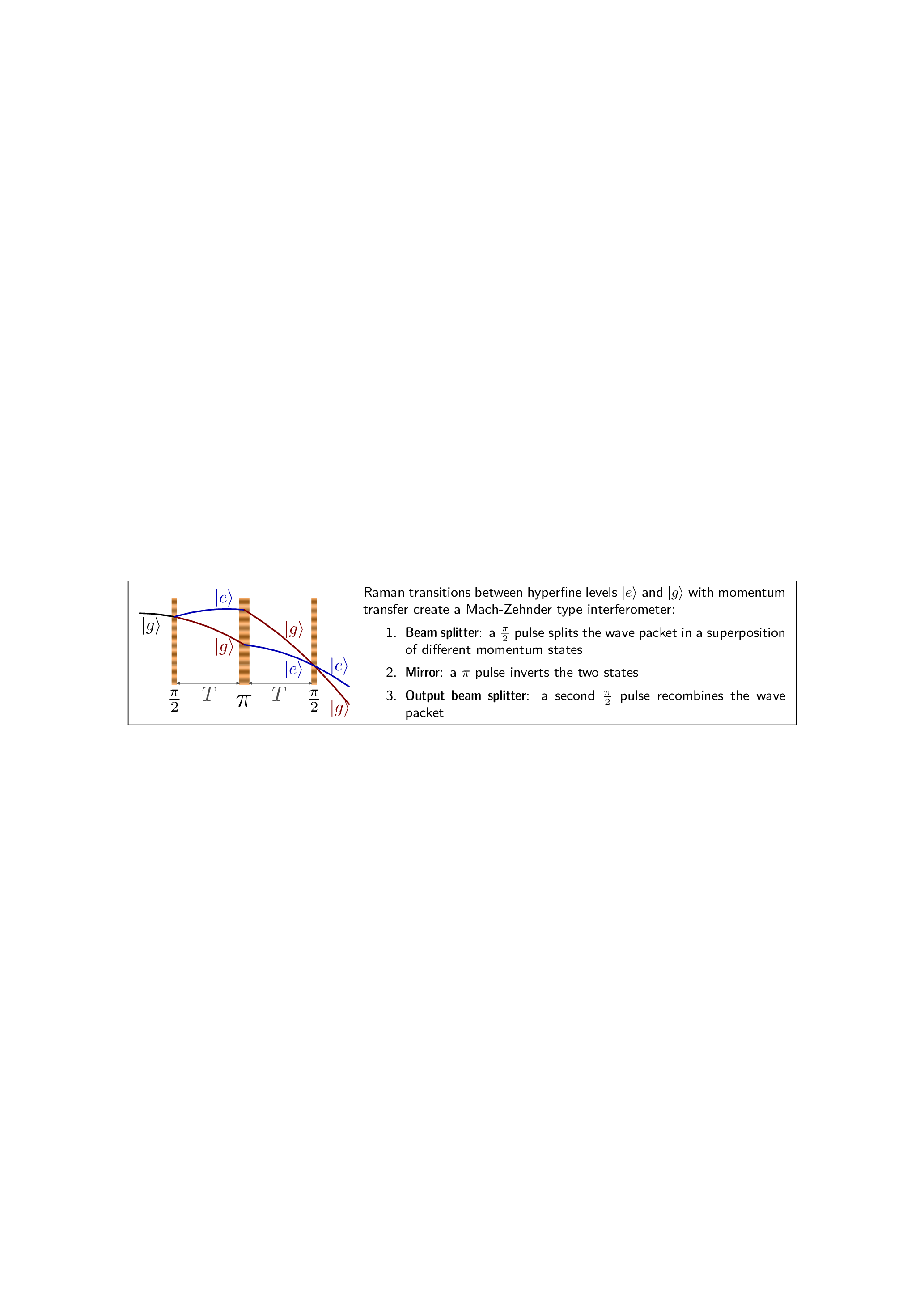}
\vspace*{-2ex}
\end{center}

\caption{Light-pulse interferometer \label{fig:light_pulse}}
\end{figure}

To detect inertial forces, an atom-interferometer must have physically
separated paths. Therefore, unlike in atomic clocks, the beam-splitting
processes must communicate momentum to the atoms. A common scheme uses
two-photons Raman transitions to coherently transfer momentum from lasers
beams to atoms \cite{Kasevich:1991}. Two hyperfine levels of an atom can
be coupled via two counter-propagating laser beams. Raman transitions
contribute two photon momenta and can be used as mirrors and
beam-splitters (see figure \ref{fig:light_pulse}).

The phase shift at the output of such a light-pulse interferometers arises from three
contributions\cite{Borde:2001}:
\begin{compactenum}
    \item the difference in the action integral along each path, 

    \item the difference in the phases imprinted on the atom waves by the beam 
	splitters and mirrors,

    \item the term due to the splitting of the wave-packets at the output
	of the interferometer. 
\end{compactenum} 
Gravito-inertial effects are found in the first term, but can also be
hidden in the second term depending on the chosen reference frame, through
movement of the Raman beams. 
When performing the calculation in an inertial frame, terms 1 and 3
nearly cancel out and leave only the term 2, the laser phase difference
between mid-points of classical (non-quantum) trajectories. This shows
that interferometric process can be pictured\cite{Peters:2001} as
measuring the position of freely falling atoms on an optical ruler made
by the two Raman lasers beating together with a wavevector given by the
difference of their wavevectors $k_\text{eff} = k_2 - k_1$. This simple
picture allows to understand the stringent requirement on the lasers for
a high-precision measurement, 
as their spectral quality determines the quality of the optical ruler.

For a light-pulse interferometer with equal times $T$ between the three
pulses, in the case of an accelerated frame with no rotation or gravity
gradient, the final phase shift is given by 
\begin{equation}
\Delta\phi = \overrightarrow{k}_\text{eff}\cdot \overrightarrow{a}\,T^2,
\label{EQ_G}
\end{equation}
where $\overrightarrow{a}$ is the local acceleration: the atom
interferometer acts as a gravimeter. Current state-of-the-art
atom-gravimeters\cite{Peters:2001} have a shot-to-shot accuracy of
$10^{-7}\,\text{m}\cdot\text{s}^{-2}$ due to technical noise.

\subsection{Micro-gravity, the route to enhanced atom-interferometric sensors}


Equation \ref{EQ_G} shows that the sensitivity of an atom-interferometer increases with interrogation
time. The longer the time the atoms spend between the beam splitters, the
greater the scaling factor between the accumulated phase shifts and the
effect they probe. In order to avoid uncontrolled residual phase-shifts,
it is best not to apply fields other than that which is probed during the
phase accumulation period. In the case of inertial sensing this implies
that atoms must be in free fall between the beam-splitting processes. In
Earth-based interferometers, with cold-atom sources, the expansion of the
atomic cloud is small, and the interrogation time is limited to a
fraction of second by the available fall height, limiting their
precision. Atom interferometry in micro-gravity allows for longer
free fall and thus increased precision.

\section{Atom-interferometric sensors in the Zero-G Airbus}

We are conducting atom interferometry experiments for inertial sensing on
board an airplane during ballistic flights. Microgravity is obtained via
20 second-long parabolas by steering the plane to cancel drag and follow
gravity. Residual acceleration is on the order of $10^{-3}\,g$. Even
though this tropospheric microgravity facility does not provide the
environmental quality of a space-borne mission, either on the ISS or on a
dedicated platform, it offers the possibility to perform test and
qualification campaigns for future space atomic inertial sensor missions.
It also provides the required environment for the first comparison of
sensors performances and possibly the first fundamental physics test with
atomic sensors in microgravity. 

\subsection{Airborne test of the equivalence principle: an Airbus as an
Einstein elevator.}

The Einstein equivalence principle states that physics in a freely
falling reference frame, in a gravitational field, is locally equivalent
to physics without any gravito-inertial fields. An atom interrogated
during its free fall in an interferometer on Earth behaves like an atom
interrogated in deep-space. But inertial-sensing interferometers have a
non-zero physical size and can be subject to tidal
effects\cite{Angonin:2006} (e.g. Lense-Thirring). An experiment carried out
nearby a massive object is therefore not equivalent to a deep space
experiment. On the other hand, there is no difference between an
experiment carried in a freely falling airplane and one on a satellite
orbiting around the Earth or the Sun.

\begin{figure}
\begin{minipage}{0.6\linewidth}
\includegraphics[width=\linewidth]{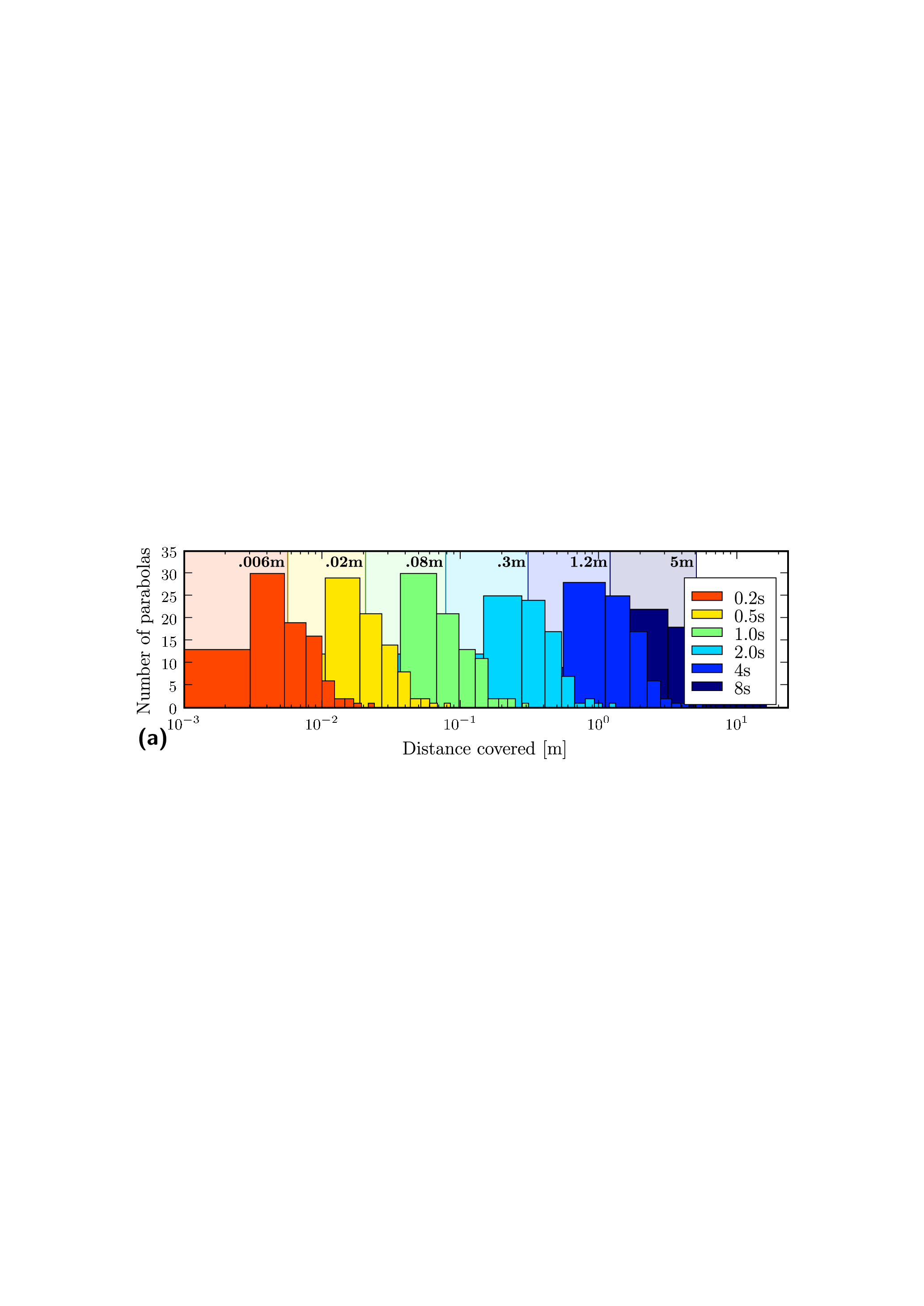}
\end{minipage}
\begin{minipage}{0.4\linewidth}
\includegraphics[width=\linewidth]{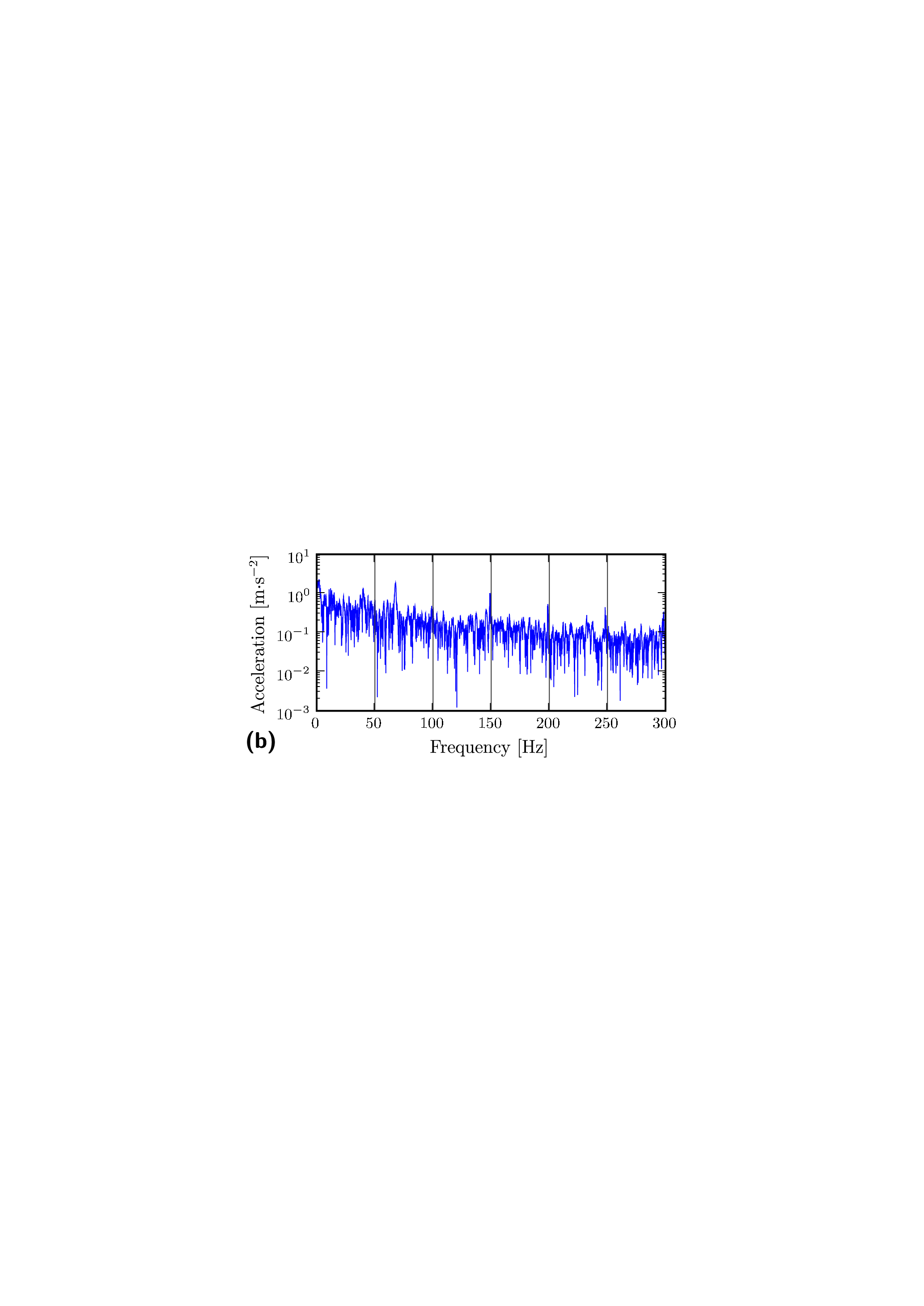}
\end{minipage}

\caption{
{\bfseries\sffamily (a)} Histograms of the displacements of a free-flying 
object, in the Airbus frame, for different times of flight. The median
value is displayed for each flight time. The displacements depend on
atmospheric conditions and vary from flight to flight. These
measurements where taken during three flights (90 parabolas), during
poor weather conditions.
{\bfseries\sffamily (b)} Acceleration noise spectrum measured on the
optical breadboard during a parabola.
}
\label{fig:acceleration}
\end{figure}

In the airplane, atoms falling in the interferometer's vacuum chamber
experience true free fall as long as they do not hit the chamber walls.
The interferometer itself is attached to the airplane and subject to
acceleration noise. All movements of the mirrors used for the
optical-ruler (the Raman lasers) strongly degrade the performance of the
interferometer\cite{Nyman:2006} (see figure \ref{fig:acceleration} (b)).
To improve the performance, the interferometer can be left freely flying,
released from the airplane, to allow for strong vibration-noise
rejection, and increased interrogation time. Since during a few seconds
of free fall an object can move by several meters in the airplane (see
figure \ref{fig:acceleration} (a)) a rigid construction, called a free
flyer, will be required to restrict and damp the displacements of the
interferometer. Choosing a free flyer with $1.2\,\text{m}$ travel gives a
50\% probability of success for a $4\,\text{s}$ long free fall for each
release of the interferometer. As the airplane is rotating about the axis
of its wings during the parabola, the rotation of the interferometer
cannot be controlled and will give rise to Sagnac shifts. They can be
canceled by doing two measurements with the area enclosed by the
interferometer reversed.

\begin{figure}[bt]

\begin{center}
\begin{minipage}{0.496\textwidth}
\includegraphics[width=\linewidth]{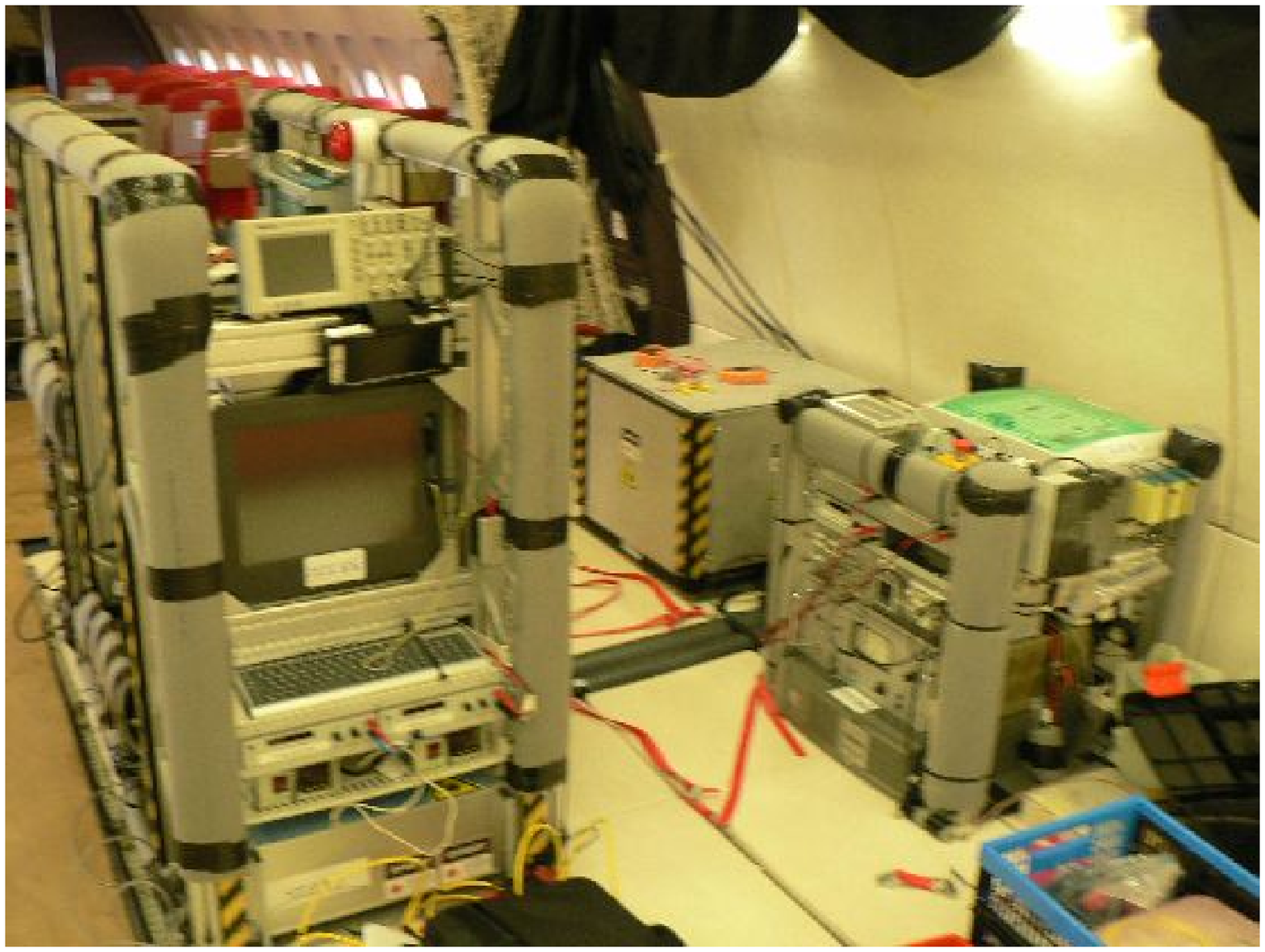}
\end{minipage}
\hfill
\begin{minipage}{0.496\textwidth}
\includegraphics[width=\linewidth]{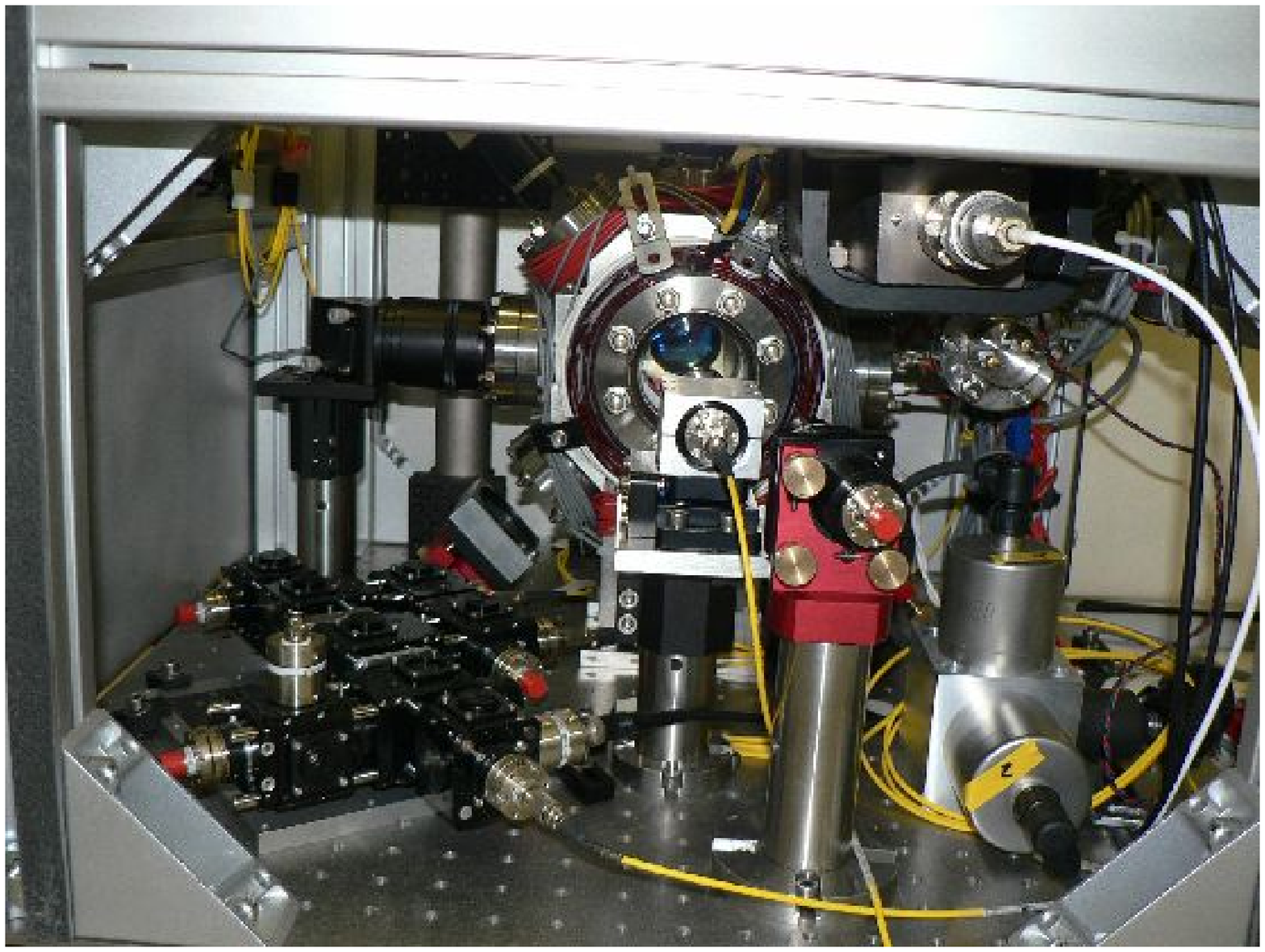}
\end{minipage}
\end{center}
\setlength{\unitlength}{\linewidth}
\begin{picture}(0,0)
\put(0.013,0.02){\sffamily\bfseries (a)}
\put(0.537,0.02){\sffamily\bfseries (b)}
\end{picture}
\vskip -2ex

\caption{
{\sffamily\bfseries (a)} 
The atom interferometer assembled in the Airbus. The main
rack, on the left, houses laser sources and control electronics. The rack
on the front right houses the uninterruptable power-supply and high-power
electrical components. The rack in the back is the atom-optics part of the
experiment.
{\sffamily\bfseries (b)}
Inside the atom-optics rack: the vacuum chamber and the free-space
optics. On the front left, the fiber beam-splitter is clamped on the
breadboard.
}
\label{fig:photos}
\end{figure}

\subsection{Test flight of a prototype micro-gravity
interferometer}

An atom interferometer is made of: a vacuum chambers with optics; lasers
sources for cooling and coherent manipulation of atoms; and a stable
oscillator (in our case a hyperfrequency source\cite{Nyman:2006}) which
serves as a frequency reference for the Raman lasers. A lab experiment
weighs at least one tonne and requires a highly-controlled environment. We
assembled a prototype atomic source suitable for inertial-sensing in a
jet airplane from the I.C.E. collaboration components\cite{Nyman:2006} (see
figure \ref{fig:photos} (a)). We prepare clouds of cold $^{87}$Rb in a
Magneto-Optical Trap (MOT) and release them for interrogation during
their free fall.

\paragraph{Novel integrated fibered source for rubidium laser-cooling.}

Laser cooling and manipulation of atoms requires frequency-stable,
narrow-linewidth laser sources. The laser systems generally used in
ultracold atom experiments are neither transportable nor reliable and
robust enough for our application. Indeed, free-space optical benches with
macroscopic cavities often need regular re-alignment. Moving away from the
standard semiconductor-laser based design\cite{Laurent:2006}, we have created
laser sources at $780\,\text{nm}$, suitable for atom interferometry with
$^{87}$Rb, using frequency-doubled fiber lasers and other telecom components at
$1560\,\text{nm}$. These novel laser sources have been described in
length elsewhere\cite{Lienhart:2007}, we will limit ourselves to outlining the
successful design choices in light of the test flight.

\begin{figure}[bt]
        \begin{minipage}{0.65\linewidth}
		\hskip -0.5ex
                \includegraphics[width=10.7cm]{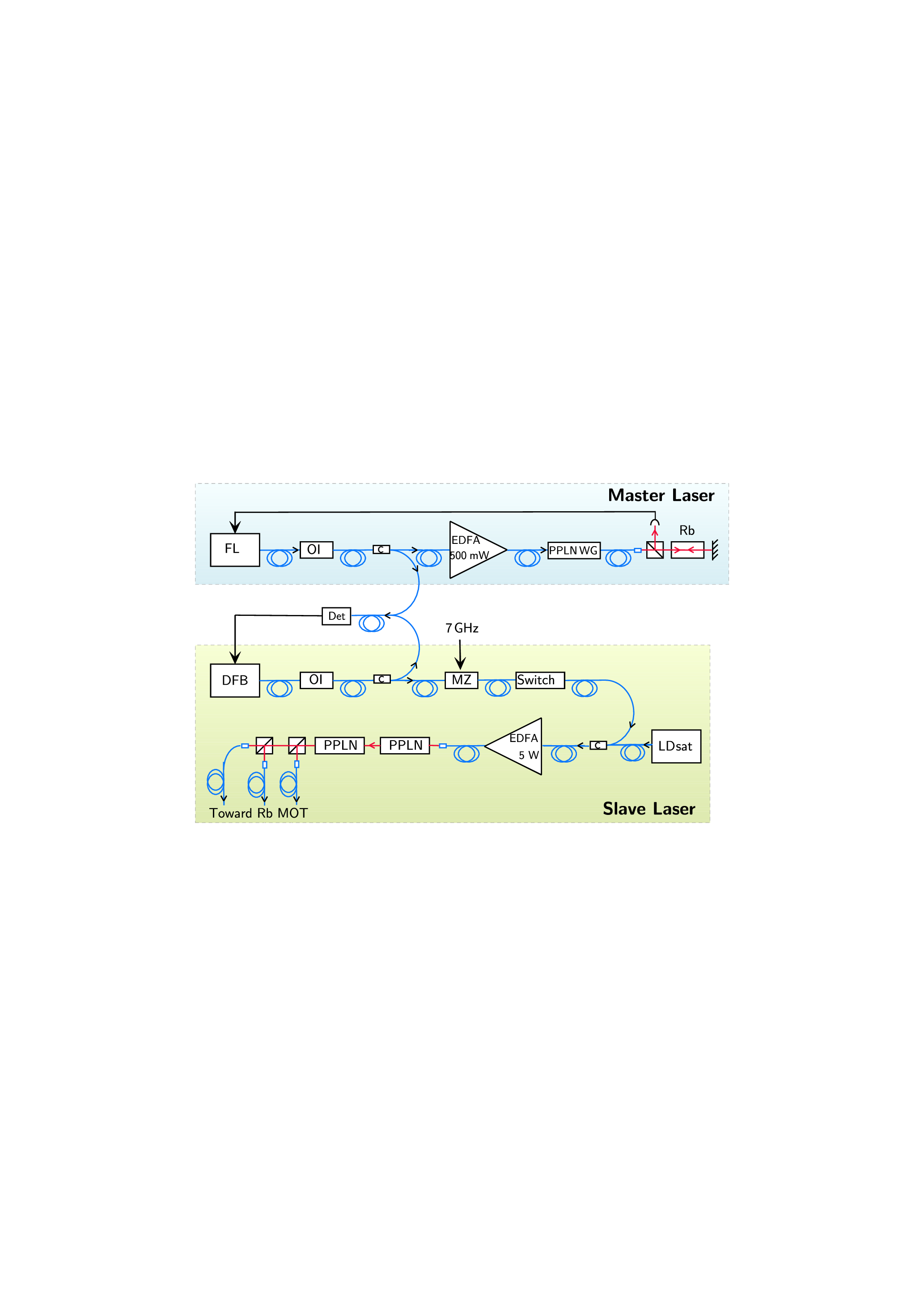}
	\end{minipage}
	\begin{minipage}{0.3\linewidth}
	{}\vskip 4.5em
	\scriptsize\sf
	\begin{tabular}{|rp{3.2cm}|}
	    \hline
	    \bfseries OI	& Optical Isolator \\
	    \bfseries FL	& Fiber Laser \\
	    \bfseries EDFA	& Erbium-Doped Fiber Amplifier \\
	    \bfseries c		& fiber coupler \\
	    \bfseries LDsat	& saturation Laser Diode \\
	    \bfseries MZ	& Mach-Zehnder modulator \\
	    \bfseries PPLN	& Periodically Poled Lithium Niobate \\
	    \bfseries PPLN WG	& PPLN Wave Guide \\
	    \bfseries DFB	& Distributed-FeedBack laser diode \\
	    \hline
	\end{tabular}
	\end{minipage}
        \caption{Diagram of our laser system. The master laser is
pictured on top, the slave below.}
        \label{montage_laser}

\end{figure}

To achieve a frequency-agile configuration, we use a master laser locked
on a rubidium transition and slave lasers which are frequency-locked to
the master laser with an arbitrary frequency difference (see figure
\ref{montage_laser}). The master laser (linewidth of $10\,\text{kHz}$) is
a monolithic semiconductor element: a $1560\,\text{nm}$ Distributed 
Feed-Back (DFB) fiber laser, amplified in a $500\,\text{mW}$ Erbium-doped
fiber amplifier and frequency doubled in a PPLN waveguide. The resulting
$780\,\text{nm}$ light is then sent into a saturated-absorption
spectroscopy setup for frequency locking to a rubidium transition. An
error signal is obtained by modulating the frequency of the master laser
for phase-sensitive detection. Control of the frequency is achieved via a
piezoelectric transducer (acting on the DFB laser) but we also change
the temperature of the DFB fiber when the piezoelectric voltage
approaches its maximum range.

The slave lasers are $80\,\text{mW}$ $1560\,\text{nm}$ DFB laser diodes
(linewidth of 1.1 MHz). After amplification through an Erbium-doped fiber
amplifier they are frequency doubled in free space with two
$2\,\text{cm}$ bulk PPLN crystals in cascade (similar
to\cite{Thompson:2003}). With a 5W fiber amplifier, we obtain $\sim
0.3\,\text{W}$ at $780\,\text{nm}$. The slave lasers are frequency-locked
to the master laser by measuring the frequency of a beat-note between the
two 1560nm lasers recorded on a fibered fast photodiode. Control of the
frequency of the slave lasers is achieved via feedback to their supply
current.

The power of the cooling laser can be adjusted by switching off the
$1560\,\text{nm}$ input laser of the fibered amplifier with an optical
switch after saturating it with a laser source at $1556\,\text{nm}$. The
$1556\,\text{nm}$ light is not frequency doubled by the PPLN crystals and
is filtered by the single-mode $780\,\text{nm}$ fibers. A very good
extinction is obtained, limited by the amplified spontaneous emission of
the fibered amplifier that is frequency-doubled. Mechanical shutters are
used to completely extinguish the lasers over long timescales (they have
a $7\,\text{ms}$ dead time), but the use of the saturation diode allows
for quicker switching times ($\sim 50\,\mu\text{s}$).

In order to laser cool $^{87}$Rb, an additional frequency (called the
repumping laser), located $7\,\text{GHz}$ away from the cooling laser, is
required. Instead of using another laser, we use a $1560\,\text{nm}$
fiber Mach-Zehnder modulator to generate two sidebands $7\,\text{GHz}$
apart. One sideband is for repumping and the other is off-resonnace, so
causes no ill effects.
 
Laser light is transported to the vacuum chamber using
polarization-maintaining optical fibers. A fiber beam-splitter (Sch\"after
and Kirchhoff) based on miniature polarizing optics divides the
laser-cooling fiber in three to provide separate beams for the operation
of the MOT.

The laser sources have proved remarkably robust during the test flight,
surviving pressure changes of $200\,\text{mPa}$, temperature changes of
$15^\circ\text{C}$, and remaining frequency-locked in spite of the noisy
environment. It is worth noting that an amplifier was damaged during
flight operations, its output power dropping by a factor of 10. This
failure did not prevent the MOT from functioning. The cause is currently
being investigated.

\paragraph{A robust transportable interferometer setup.}

Keeping with the philosophy of a flexible prototype, the atomic-physics
part of the interferometer was built using standard lab equipment mounted
on a $600\times600\,\text{mm}$ optics breadboard (see figure
\ref{fig:photos} (b)). A rigid frame is bolted through the breadboard,
holding the vacuum chamber to protect it and meet flight security
requirements.

The core of the apparatus is a stainless-steel ultra-high vacuum chamber
in which the laser beams intersect for trapping and manipulating the
atoms. The chamber has two $63\,\text{mm}$ diameters windows and eight
lateral $40\,\text{mm}$ ports. The large number of available ports allows
us to dedicate separate windows for different beams and for observation.
MOT and compensation coils are directly wound onto the chamber. The
rubidium atoms are released from commercially-available alkali-metal
dispensers directly into the MOT chamber. While operating the
interferometer, the dispensers are run continuously and a dilute ($<
10^{-8}\,\text{mBar}$) rubidium vapor fills the chamber. There are two
pumps: an ion pump and a getter pump, maintaining the required vacuum
even during night power cut. No special care has been taken to ensure
that all parts of the system are non-magnetic and the ion pump was not
shielded. We relied solely on the compensation coils to cancel out the
magnetic fields in the surroundings of the atoms.

The six counter-propagating lasers beams of the MOT are made of three
retro-reflected beams each expanded out of a fiber by an out-coupler
producing a $25\,\text{mm}$ diameter beam. The couplers are positioned on
kinematic mounts (New Focus 9071) held by $38\,\text{mm}$ posts bolted on
the breadboard. The setup was optimized in our lab in Palaiseau, then
carried on a truck $500\,\text{km}$ away to Bordeaux and loaded into the
airplane with no particular precautions. Every day temperature cycled
from $6^\circ\text{C}$ to $20^\circ\text{C}$. We did not notice any
misalignment.

The laser source driving the Raman transition is similar to the cooling
laser. The second Raman frequency is achieved as with the repumping
frequency by intensity-modulating the laser light. The presence of the
second sideband adds new paths to the interferometer, but they need not
be taken in account as these secondary interferometers are not closed.
For the Raman pulse manipulation, MOT coils are switched from quadrupole
configuration to dipole configuration to provide a polarizing field
raising the degeneracy between Zeeman sub-levels. The intensity of the
lasers can be up to 20 times the saturation intensity of rubidium, which
allows for short Raman pulses with weak velocity selection to address
broad momentum distributions. The pulse is controlled via an
acousto-optical modulator after the frequency-doubling stage. Raman
transfer was not tested during this first flight.

\begin{figure}
\begin{center}
    \includegraphics[width=\linewidth]{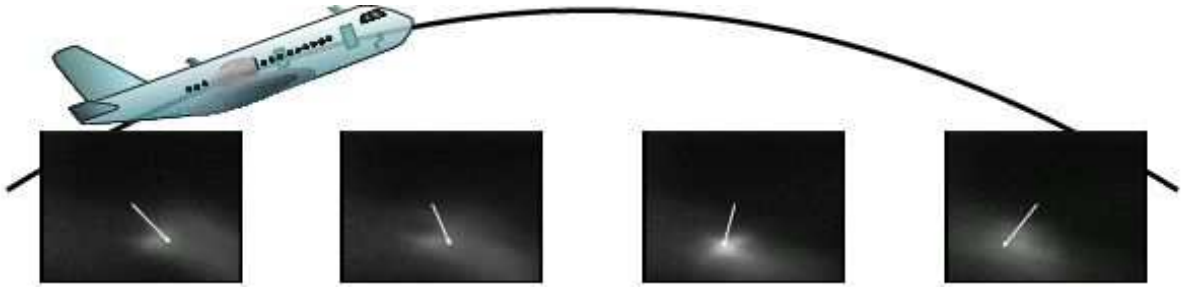}
\end{center}

\caption{Images of molasses at different airplane angles. The tilt in the
Earth's magnetic field produces an imbalance in the radiation pressure
during the molasses phase, and alters the direction in which the atoms
escape\label{fig:tof}. The arrows connect the positions of the initial
trapped cloud to the escaping atoms. The escape direction
does not directly relate to the pitch angle of the airplane, as its
bearing also changes the direction of the magnetic-field.}

\end{figure}

We load $~10^9$ atoms of $^{87}$Rb in the MOT in one second. A photodiode
monitors the fluorescence, which is proportional to the number of atoms.
We release the atoms from the MOT and further cool them through a brief
phase of optical molasses during which we can prepare the atoms in the
lower hyperfine state by turning off the repumping light. The light
pulses for the interferometer (see figure \ref{fig:light_pulse}) can then
be applied. The cooling laser is turned back on to detect the atoms still
in the MOT volume in a state selective way.
The flight of the atoms in the optical molasses was recorded during the
test flight. Our preliminary setup had no magnetic shield and the
rotation of the Earth's magnetic field created uncompensated Zeeman
shifts. These shifts imbalance the radiation pressure during molasses,
limiting the atomic escape velocity (see figure \ref{fig:tof}).

\section{I.C.E.: the next-generation apparatus}


Temperatures achieved through laser-cooling techniques reach a limit of
around $1\,\mu\text{K}$
($v_\text{rms}\sim1\,\text{cm}\cdot\text{s}^{-1}$). Ballistic expansion
of laser-cooled atoms thus limits interrogation times to a few seconds
\cite{Lecoq:2006}. To make full use of micro-gravity (e.g. with a
space-borne experiment) further cooling is needed. The limiting factor
for atom-interferometric metrology is the size of the atomic cloud after
expansion, given by both its initial size and velocity spread. This size
is closely related to the phase-space density of the source, which can be
seen as an atom-optic equivalent of the luminance of a photonic source.
The source of maximum luminance in optics is the laser, and which is
widely used in photon-interferometry. In atom-optics such a source is a
Bose-Einstein condensate. The intrinsic linewidth of a Bose-Einstein
condensate is very narrow, however the momentum-width of a freely falling
condensate depends on the trap release process. 

Within the I.C.E. project\cite{Nyman:2006}, we are building a
next-generation atom interferometer making use of quantum degenerate
atomic gases for increased interrogation times during ballistic flights.
Degenerate atom sources can be achieved by evaporative cooling, where the
depth of a non-dissipative trap is lowered to eject higher energy atoms
while relying on two-body collisions to thermalize the cloud, thus
lowering its temperature. Magnetic traps, where atoms are trapped near a
minimum of magnetic field, are most often used to perform this
evaporation. However, releasing atoms from a magnetic trap is an
ill-controlled process that will affect the sensor performances.
On the contrary, optical-dipole traps, where atoms are trapped near a
maximum of laser light intensity, can be controlled with much more
precision. Ramping down to zero the power of the trapping laser yields a
jerk-free release process. In the I.C.E. interferometer, we transfer the
atoms from the MOT to an optical-dipole trap made of two intersecting
laser beams at $1565\,\text{nm}$. Efficient capture of the atoms is
achieved for a trap depth of a few times their kinetic energy. The
phase-space volume of a dipole trap is limited by the available laser
power. As a MOT is a rather large cloud, it is best to capture it with
the spatially largest possible trap while still deep enough to load the
atoms. Using a $50\,\text{W}$ fiber laser, we can load the atoms using
trap diameters up to $350\,\mu\text{m}$.

To permit a short duty cycle that allows for high precision measurement
and high accuracy, the production of the degenerate atomic source must be
fast. For the evaporation process to be efficient, the collision rate
needs to increase as the number of trapped atoms decreases. However,
lowering the power of the trapping laser reduces the depth of the trap,
but not its size, and, as the trap depth goes to zero, so does the
density of the trapped atoms and the collision rate. We use a mechanical
zoom to change the diameter of the trapping laser beam during the
evaporation and optimize the collision rate for quick evaporative
cooling\cite{Kinoshita:2005}. 


Bose-Einstein condensates are relatively high-density samples, and the
interactions between atoms cannot be neglected. These interactions give
rise to uncontrolled shifts in
interferometers\cite{Gupta:2003,Lecoq:2006}. Inertial-sensing atom
interferometers have a physical extent, and the wave-packets in the
different arms of the interferometers spend most of their time
non-overlapping. The collisional shifts occur at the beam-splitters,
where each atom interacts with the atoms in the other arm
\cite{Lecoq:2006}. On the contrary, due to Pauli blocking, ultracold,
spin-polarized, degenerate fermions do not interact. They suffer no
interaction shifts \cite{Gupta:2003}, but, as their phase-space density
cannot exceed unity, they form less bright sources and, when released
from tight traps, their increased momentum spectrum could limit
interrogation times.
However, Pauli blocking in Fermi gases and interactions in Bose gases
yield similar orders of magnitude\cite{TrapRelease} for momentum
broadening over the range of accessible experimental parameters as long
as the interaction is not suppressed via magnetically-tunable Feshbach
resonances.
Nevertheless, both collisional-shifts and momentum broadening can be
strongly reduced by adiabatically opening the trap and reducing the
density before the release, or by the use of an external magnetic field
to tune the interactions (Feshbach resonances). The I.C.E. interferometer
has been designed to operate with fermionic $^{40}$K in addition to
bosonic $^{87}$Rb to allow for comparison of the achievable precisions
using different species.

\section*{Conclusion}

We have successfully tested a cold atom source for inertial sensing in
aircraft parabolic flights. With the use of a free-flyer, it will allow
interrogation of the freely-falling atoms during several seconds, paving
the way for high-precision inertial sensors. The interrogation time is
then limited be the residual acceleration. The aircraft is a valid frame
for fundamental tests of gravitation theories. Our preliminary results
show that laboratory experiments can be adapted for this new experimental
platform. Unlike orbital platforms, development cycles on ground-based
facilities, either on the plane (it took us 3 months to assemble the test
prototype), or on drop towers\cite{Vogel:2006}, can be sufficiently
short to allow for the rapid technological evolution for future sensors.
Indeed high-precision drag-free space-borne applications require further
progress on achieving longer interrogation times using ultra-low velocity
atoms. Our new-generation degenerate atomic source design minimizes
trap-release and interaction energies for these purposes.

\section*{Acknowledgments}

The I.C.E. collaboration is funded by the CNES, as are JPB, RAN and NZ's
salaries. Further support comes from the European Union STREP consortium
FINAQS. Laboratoire Charles Fabry, ONERA, and LNE-SYRTE are affiliated to
IFRAF\cite{IFRAF}.



\begin{thebibliography}{99}
\bibitem{TrapRelease}
When suddenly releasing interacting Bose gases from a trap, the release
momentum is given by the energy of the bosons in the trap. For trapping
frequencies of $\omega$, the harmonic oscillator length related to the
trap is $a_\text{ho} = \sqrt{\frac{\hbar}{m\,\omega}}$, and the release
wave-vector is $k_\text{release} \sim \frac{1}{a_\text{ho}}
(15\,N\frac{a}{a_\text{ho}})^{\slantfrac{1}{5}}$, with $a$ the
scattering length of the atoms. For Fermi gases, the spread in
wave-vector is given by $k_\text{F} \sim \frac{1}{a_\text{ho}}
(48\,N)^{\slantfrac{1}{6}}$.
\end{thebibliography}
\end{document}